# Impurity scattering effects on the low-temperature specific heat of $d$-wave superconductors

C. F. Chang[1], J.-Y. Lin[2,*], and H. D. Yang[1]

[1]*Department of Physics, National Sun Yat-Sen University, Kaohsiung 804, Taiwan ROC*

[2]*Institute of Physics, National Chiao Tung University, Hsinchu 300, Taiwan ROC*

Very recently impurity scattering effects on quasiparticles in $d$-wave superconductors have attracted much attention. Especially, the thermodynamic properties in magnetic fields $H$ are of interest. We have measured the low-temperature specific heat $C(T,H)$ of $La_{1.78}Sr_{0.22}Cu_{1-x}Ni_xO_4$. For the first time, the impurity scattering effects on $C(T,H)$ of cuprate superconductors were clearly observed, and are compared with theory of $d$-wave superconductivity. It is found that impurity scattering leads to $\gamma(H)=\gamma(0)(1+D\times(H/H_{c2})\times\ln(H_{c2}/H))$ in small magnetic fields. Most amazingly, the scaling of $C(T,H)$ breaks down due to impurity scattering.

Tunneling and ARPES experiments which are sensitive to either the interface of the junction or surface of the sample have suggested a dominant $d$-wave pairing symmetry in hole-doped cuprate superconductors [1,2]. Still, low-temperature specific heat ($C$) is thought to be one of the unique experiments which provide evidence of $d$-wave pairing in bulk properties. The $T^2$ temperature dependence of the electronic term in $C$ at zero magnetic field $H=0$ and the $H^{1/2}$ dependence of the linear term coefficient $\gamma$ have been interpreted as strong evidence of the line nodes of order parameter [3-12]. Very recently, the scaling behavior of the electronic specific heat contribution $C_e(T,H)$ has been predicted by theory [13,14], and confirmed by experiments [5,9-11]. However, several papers reported that the non-linear $H$ dependence of $\gamma$ was also observed in conventional superconductors [15,16], and raised the question whether the $H^{1/2}$ dependence of $\gamma$ is indeed due to $d$-wave pairing. In addition, although most studies of $C(T,H)$ in cuprates agree on the $H^{1/2}$ dependence of $\gamma$, there remains controversies on the existence of the $T^2$ term at $H=0$. Chen et al. have presented data showing clear evidence of the $T^2$ term in $La_{1.78}Sr_{0.22}CuO_4$ and disappearance of this $T^2$ term in $H$, both consistent with the predictions for $d$-wave superconductivity [5]. Nevertheless, in some other works, evidence of the $T^2$ term was either ambiguous or had to be identified through sophisticated fit [6-10]. These difficulties make the $C(T,H)$ studies of the impurity-doped cuprate superconductors particularly of interest. If the recently developed theory [17-19] of the impurity scattering effects on the quasiparticle excitation in cuprates can be verified by $C(T,H)$ measurements, it would strongly indicate that the observed properties of $C(T,H)$ are characteristic of $d$-wave pairing. These studies may also help to improve the theories of the quasiparticles in cuprates. Furthermore, since a small impurity scattering rate can cause disappearance of the $T^2$ term, it is desirable to know the magnetic field dependence of $C(T,H)$ in the impurity-doped cuprates. Comparisons between $C(T,H)$ of the nominally clean samples and that of the impurity-doped ones may generate fruitful implication on the existing puzzles.

To serve these purposes, $La_{1.78}Sr_{0.22}Cu_{1-x}Ni_xO_4$ samples were chosen for two main reasons. $C$ of the Ni-doped samples has a much smaller magnetic contribution than that of the Zn-doped samples, and the data analysis can be simplified. Moreover, $La_{1.78}Sr_{0.22}CuO_4$ has been compellingly shown to be a clean $d$-wave superconductor [5], and is ideal to compare with the Ni-doped samples. Polycrystalline samples of $La_{1.78}Sr_{0.22}Cu_{1-x}Ni_xO_4$ with nominal $x=0$, 0.01, and 0.02 were carefully prepared from $La_2O_3$, $SrCO_3$, and $CuO$ powder of 99.999% purity. Details of the preparation were described elsewhere [5]. The powder x-ray-diffraction patterns of all samples used in the experiments show a single T phase with no detection of impurity phases. The transition temperature $T_c$ by the midpoint of the resistivity drop is 28.7, 21.2 and 17.4 K for $x=0$, 0.01, and 0.02, respectively. The transition width (90% to 10% by the resistivity drop) of $T_c$ is 3 K or less for all samples, suggesting a decent homogeneity. $C(T)$ was measured from 0.6 to 9 K with a $^3$He thermal relaxation calorimeter using the heat-pulse technique. The precision of the measurements in the temperature range is about 1%. To test the calibrations of the thermometer and the measurements in $H$, a copper sample was measured, and the scatter of data in different magnetic fields is about 3% or better. Details of the calorimeter calibrations by the copper sample can be found in Ref. [5].

The analysis of $C(T,H)$ was carried out for data from 0.6 to 7 K. Varying the temperature range to 8 K or to 6 K does not lead to any significant change of the results. Both the individual-field and global fit have been executed, and give similar results and conclusion. In this Letter, the results from the individual-field fit are reported. Data of all samples are described by

$$C(T,H)=\gamma(H)T+\beta T^3+ nC_{S=2}(T,H) \quad (1)$$

where $\beta T^3$ is the phonon contribution and $nC_{S=2}$ is the magnetic contribution of spin-2 paramagnetic centers (PC's) associated with $CuO_2$ planes [20-22]. Since $La_{1.78}Sr_{0.22}Cu_{1-x}Ni_xO_4$ has only $CuO_2$ planes and lacks CuO chains, $nC_{S=2}$ was used rather than the conventional Schottky anomaly, which was thought to be related to CuO Chains [7,22]. Phenomenologically, inclusion of $nC_{S=2}$ also yields a better fit than that of the Schottky anomaly.

$C(T,0)$ of samples with $x=0$, 0.01 and 0.02 is shown in Fig. 1. For $x=0$, at zero field $C/T$ vs. $T^2$ shows an obvious downward curve at low temperatures due to the $T^2$ term in $C$. For $x=0.01$, this downward curve becomes a straight line except below 1 K where the contribution from the magnetic contribution becomes important. An increase in $\gamma$ with increasing $x$ can also be recognized directly from data shown in Fig. 1. Both disappearance of the $T^2$ term and the increase in $\gamma$ are considered as manifestations of the impurity scattering. The low-temperature upturn in $C/T$ of both $x$=0.01 and 0.02 can be attributed to $nC_{S=2}$ by the solid lines resulted from the fit by Eq. (1). To further show the quality of the fit in $H$, $C(T,H)$ of $x$=0.01 at low temperatures is shown in Fig. 2(a) as an example, together with the solid lines representing the fit of data to Eq. (1). The results illustrate that $C(T,H)$ of $La_{1.78}Sr_{0.22}Cu_{1-x}Ni_xO_4$ can be satisfactorily described by Eq. (1). The contribution of $nC_{S=2}$ compared with other terms is shown in Fig. 3. As expected, $n$ resulting from the fit does not change significantly with $H$, however with variation in $H \geq 4$ T as shown in Fig. 2(b). Similar results of $n$ vs. $H$ can be found in all three samples. It is likely that the effective Hamiltonian for $C_{S=2}$ in Ref. [20] results from the experimental data with $H<4$ T [21], and is most suited for low magnetic fields. From the low-field fitting results, $n$ of $x$=0, 0.01, and 0.02 is about 0.3, 0.9, and $1.8 \times 10^{-4}$ respectively. The value of $n$ for $x$=0 is taken from the fit of the data in $H$ and implemented into the fit at $H$=0. The solid line for $x$=0 in Fig. 1(a) shows that the data can accommodate a small $nC_{S=2}$.

For a clean $d$-wave superconductor in a finite field $H$, an increase in $\gamma$ is predicted to be proportional to $H^{1/2}$ at low temperatures due to the Doppler shift on the quasiparticle energy [3,4]. In the unitary limit, impurity scattering leads to a modification to the density of states, and the $H$ dependence of $\gamma$ becomes [17-19]

$$\gamma(H)=\gamma(0)(1+D \times (H/H_{c2}) \times \ln(H_{c2}/H))$$

(2)

where $D \approx \Delta_0/32\Gamma$. $\Delta_0$ is the superconducting gap, $\Gamma$ is the impurity scattering rate, and $H_{c2}$ is the upper critical field. The unitary limit is widely considered as a good approximation to the nature of the impurity scattering in cuprates, and is supported by experimental evidences. To compare $\gamma(H)$ of the clean sample with that of the Ni-doped ones, $\gamma$ vs. $H^{1/2}$ of all samples was plotted in Fig. 4. If $\gamma$ has a $H^{1/2}$ dependence as expected in a clean sample, the data will follow a straight line as represented by the dash line in Fig. 4. Indeed, data of the sample with $x$=0 indicate a clear $H^{1/2}$ dependence of $\gamma$ (Fig. 4(a)). In Ni-doped samples, the $H$ dependence of $\gamma$ is smaller than in the clean one, and the data show a pronounced curvature for small $H$ (Fig. 4(b) and (c)). This behavior makes $\gamma(H)$ of Ni-doped samples distinct from that of the clean one. Thus the effect of the impurity scattering is distinguished. Actually, $\gamma(H)$ of both Ni-doped samples can be well described by Eq. (2) with reasonable parameters as shown by the solid line in Fig. 4(b) and (c). The fit gives $\Gamma/\Delta_0$=0.020 and 0.025 for $x$=0.01 and 0.02, respectively, with $H_{c2} \approx 38$ T. A twice increase in $\Gamma/\Delta_0$ is expected for $x$=0.02 by the nominal doping concentration; nevertheless, this small increase in $\Gamma/\Delta_0$ is in accord with a less rapid $T_c$ suppression in the $x$=0.02 sample. Furthermore, as a result of the impurity scattering, the values of $\gamma/\gamma_n$ corresponding to those of $\Gamma/\Delta_0$ are in good agreement with the calculated values in Refs. [17,18] for both Ni-doped samples. On the other hand, a try to fit $\gamma(H)$ of the clean sample by Eq. (2) has proven to be fruitless and resulted in a unrealistic $H_{c2}$>1000 T.

The most crucial test of the recent theory for a d-wave superconductor with impurities probably lie on the breakdown of the scaling behavior of $C_e(T,H) \equiv C(T,H) - \gamma(H=0)T - \beta T^3 - nC_{S=2}$.

For a clean $d$-wave superconductor, if $C_e/(TH^{1/2})$ vs. $H^{1/2}/T$ is plotted, all data at various $T$ and $H$ should collapse into one scaling line according to the recent scaling theory [13,14]. This scaling of $C_e(T,H)$ has been observed in $YBa_2Cu_3O_{7-\delta}$ and $La_{1-x}Sr_xCuO_4$ samples [5,9-11]. As shown in Fig.5 (a), $C_e(T,H)$ of $La_{1.78}Sr_{0.22}CuO_4$ follows this scaling. However, a recent theory predicts that strong impurity scattering can cause breakdown of the scaling [17,23]. This dramatic effect is best illustrated in Fig. 5(b) and (c). In contrast to the scaling of $C_e(T,H)$ of the clean sample, $C_e(T,H)$ data of Ni-doped samples split into individual isothermal lines as predicted by the numerical calculations [17].

The very theory also suggests that Eq. (2) is exact only in fields $H \ll H^*$ where $H^*/H_{c2} \approx \Gamma/\Delta_0$ [17,18]. Though, $\gamma(H)$ should not deviate Eq. (2) too much if $H$ is slightly larger than $H^*$ [24]. In case of $H \gg H^*$, $\gamma(H)$ would mimic the $H^{1/2}$ behavior [18]. With $H^* \approx 1$ T in the present experiments, $\gamma(H)$ in Fig. 4(b) and (c) behaves exactly like what is expected. In small $H$, the weak magnetic field dependence is well described by Eq. (2). In large $H$, the data do not obey Eq. (1) as well as in small $H$, and a distinction between Eq. (2) and the $H^{1/2}$ dependence is less easily made. Therefore, the less satisfactory fit in high fields merely reflects the limit of Eq. (2) as expected from the theory.

It is noted that $n$ of the spin-2 PC's increases with the doping concentration $x$. However, it is unlikely that the magnetic contribution in $C(T,H)$ comes directly from the Ni ions since $n$ is two order of magnitude smaller than $x$. Recently, it has been reported that the nominal magnetic Ni ions do not disturb the spin correlation in $CuO_2$ planes even on Ni sites at small $x$ in overdoped cuprates [25]. In both $C$ and susceptibility $\chi$ measurements, no paramagnetic contribution from Ni was observed. $C$ reported in this Letter and the related preliminary studies on $\chi$ are consistent with these results [26]. The larger $nC_{S=2}$ in the Ni-doped samples probably comes from the defects in $CuO_2$ planes, which are induced by Ni substitution. On the other hand, Zn substitution has strong effects on $C$ (and $\chi$). The large magnetic contribution usually makes the studies of the impurity scattering effects on $C(T,H)$ inconclusive [27,28]. More detailed studies on these novel properties of $C$ and $\chi$ in Ni- or Zn-doped cuprates are desirable.

In conclusion, the impurity scattering effects on $C(T,H)$ of $d$-wave superconductors have been clearly identified. The weak $H$ dependence of $\gamma(H)$ in small magnetic fields and the breakdown of the scaling behavior of $C_e(T,H)$ both are consistent with predictions of the present theory. It is thus suggested that the unconventional features observed in $C(T,H)$ of either clean or impurity-doped cuprate superconductors are intrinsic bulk properties of $d$-wave superconductivity.

We would like to thank H. F. Meng for indispensable discussions, and Y. H. Tang for technical assistance. This work was supported by National Science Council of Republic of China under Contract Nos. NSC89-2112-M-110-008 and NSC89-2112-M-009-007.

**References**

[*]Author to whom correspondence should be sent

[1] J. F. Annett, N. D. Goldenfeld, and A. J. Leggett, in *Physical Properties of High Temperature Superconductors* **V**, edited by D. M. Ginsberg, World Scientific, Singapore (1996).

[2] H. Ding, M. R. Norman, T. Yokoya, T. Takeuchi, M. Randeria, J. C. Campuzano, T. Takahashi, T. Mochiku, and Kadowaki, Phys. Rev. Lett. **78**, 2628 (1997) and references therein.

[3] G. E. Volovik, JETP Lett. **58**, 469 (1993).

[4] H. Won and K. Maki, Europhys. Lett. **73**, 2744 (1995).

[5] S. J. Chen, C. F. Chang, H. L. Tsay, H. D. Yang, and J.-Y. Lin, Phys. Rev. B **58**, R14 753 (1998).

[6] K. A. Moler, D. J. Baar, J. S. Urbach, Ruixing Liang, W. N. Hardy, and A. Kapitulnik, Phys. Rev. Lett. **73**, 2744 (1994).

[7] K. A. Moler, D. L. Sisson, J. S. Urbach, M. R. Beasley, and A. Kapitulnik, Phys. Rev. B **55**, 3954 (1997).

[8] R. A. Fisher, J. E. Gordon, S. F. Reklis, D. A. Wright, J. P. Emerson, B. F. Woodfield, E. M. McCarron III, and N. E. Phillips, Physica C **252**, 237 (1995).

[9] D. A. Wright, J. P. Emerson, B. F. Woodfield, J. E. Golden, R.A. Fisher, and N. E. Phillips, Phys. Rev. Lett. **82**, 1550 (1999).

[10] R. A. Fisher, B. Buffeteau, R. Calemezuk, K. W. Dennis, T. E. Hargreaves, C. Marcenat, R. W. McCallum, A. S. O'Connor, N. E. Phillips, and A. Schilling, unpublished.

[11] B. Revaz, J.-Y. Genoud, A. Junod, A. Erb, and E. Walker, Phys. Rev. Lett. **80**, 3364 (1998).

[12] N. Momono and M. Ido, Physica C **264**, 311 (1996).

[13] S. A. Simon and P. A. Lee, Phys. Rev. Lett. **78**, 1548 (1997).

[14] G. E. Volovik, JETP Lett. **65**, 491 (1997).

[15] A. P. Ramirez, Phys. Lett. A **211**, 59 (1996).

[16] J. E. Sonier, M. F. Hundley, and J. W. Brill, Phys. Rev. Lett. 82, 4914 (1999).

[17] C. Kubert and P. J. Hirschfeld, Solid State Com. **105**, 459 (1998); C. Kubert and P. J. Hirschfeld, Phys. Rev. Lett. **80**, 4963 (1998).

[18] Y. S. Barash, A. A. Svidzinskii, and V. P. Mineev, JETP Lett. **65**, 638 (1997).

[19] E. Janod, R. Calemczuk, J.-Y. Henry, J. Flouquet, Physica C **281**, 176 (1997); the theory including the effects of impurity scattering was written by G. E. Volovik in Appendix A of this paper.

[20] O. N. Bakharev, M. V. Eremin, and M. A. Teplov, JETP Lett. **61**, 515 (1995).

[21] P. G. Baranov and A. G. Badalyan, Solid State Communications **85**, 987 (1993).

[22] J. P. Emerson, D. A. Wright, B. F. Woodfield, J. E. Golden, R.A. Fisher, and N. E. Phillips, Phys. Rev. Lett. **82**, 1546 (1999).

[23] G. E. Volovik (private communication).

[24] P. J. Hirschfeld (private communication).

[25] T. Nakano, N. Momono, T. Nagata, M. Oda, and M. Ido, Phys. Rev. **58**, 5831 (1998).

[26] J.-Y. Lin, H. D. Yang, T. I. Hsu, and H. C. Ku, unpublished.

[27] J.-Y. Lin, C. F. Chang, and H. D. Yang, Physica B, in press.

[28] D. L. Sisson, S. G. Doettinger, A. Kapitulnik, R. Liang, D. A. Bonn, and W. N. Hardy, cond-mat/9904131.

**Figure captions**

Fig. 1. $C/T$ vs. $T^2$ of $La_{1.78}Sr_{0.22}Cu_{1-x}Ni_xO_4$ with $x=0, 0.01$ and $0.02$ at $H=0$. The solid lines are the results of the fit by Eq. (1). Inset: $C/T$ vs. $T$ for $T<2$ K, where the contribution from the $T^2$ term is apparent.

Fig. 2. (a) $C/T$ vs. $T^2$ of $La_{1.78}Sr_{0.22}Cu_{0.99}Ni_{0.01}O_4$ in magnetic fields. The solid lines are the results of the fit by Eq. (1). For clarity, only data in $H=0, 0.2, 1, 4,$ and 8 T are shown. (b) The concentration $n$ of the spin-2 PC's from the fit.

Fig. 3. The components of $C(T,H)$ of $La_{1.78}Sr_{0.22}Cu_{0.99}Ni_{0.01}O_4$.

Fig. 4. Normalized $\gamma(H)$ vs. $H^{1/2}$ for three $La_{1.78}Sr_{0.22}Cu_{1-x}Ni_xO_4$ samples. The solid lines are the results of the fit by Eq. (2), which includes the impurity effects on $C(T,H)$. Dash lines represent $\gamma(H) \propto H^{1/2}$ expected in clean $d$-wave superconductors. In (a) no solid line is presented since the fit by Eq. (2) gives a unrealistic value of $H_{c2}>1000$ T. $\gamma_n=12$ mJ/mol K$^2$ is the normal state $\gamma$ of the samples [5].

Fig. 5. Plots of $C_e/(TH^{1/2})$ vs. $H^{1/2}/T$ for (a) $x=0$, (b) $x=0.01$, and (c) $x=0.02$. Note that the scaling which holds in (a) breaks down in (b) and (c) due to impurity scattering.

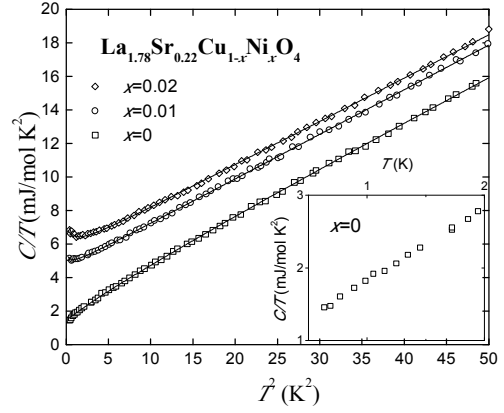

Fig. 1.

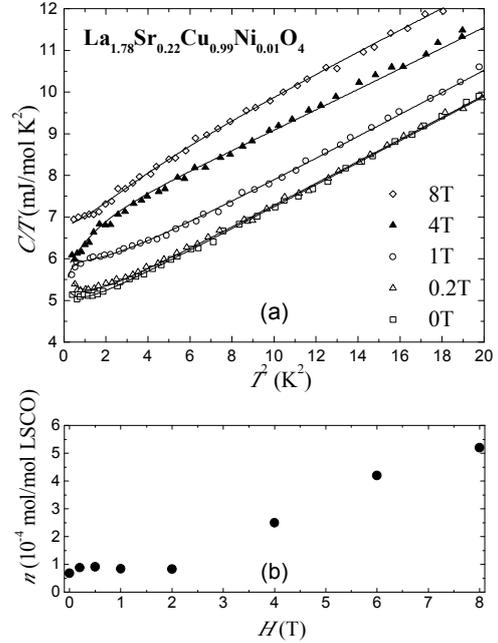

Fig. 2

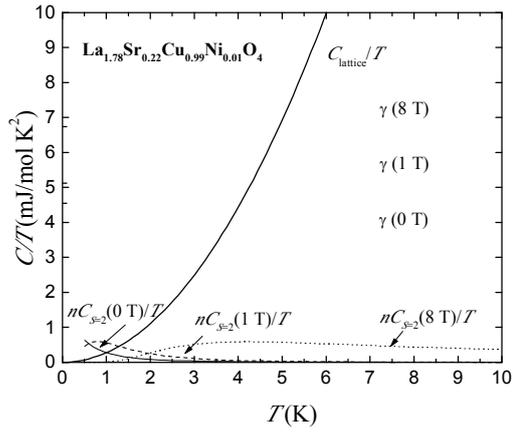

Fig. 3.

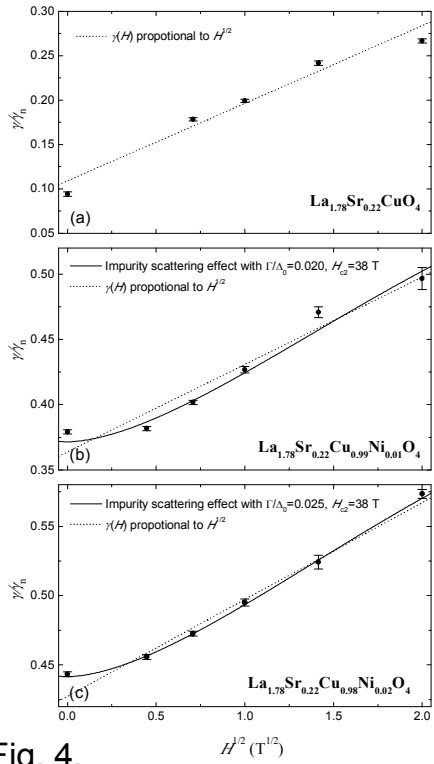

Fig. 4.

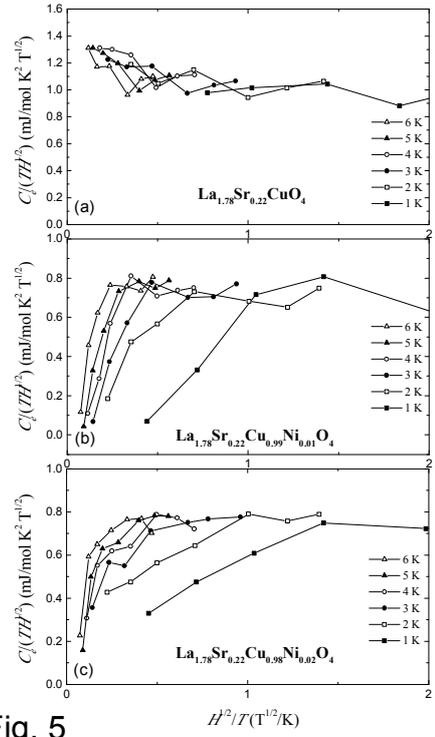

Fig. 5